\documentstyle[11pt]{article}

\begin{document}

\renewcommand{\vec}[1]{{\bf #1}}       
\def\beq{\begin{eqnarray}}    
\def\eeq{\end{eqnarray}}      

\def\tr{\,\mbox{tr}\,}                  
\def\Tr{\,\mbox{Tr}\,}                  
\def\sTr{\,\mbox{sTr}\,}                
\def\Res{\,\mbox{Res}\,}                
\renewcommand{\Re}{\,\mbox{Re}\,}       
\renewcommand{\Im}{\,\mbox{Im}\,}       
\def\lap{\Delta}                        
\def\sla{\!\!\!\slash}

\def\al{\alpha}
\def\be{\beta}
\def\ch{\chi}
\def\ga{\gamma}
\def\de{\delta}
\def\vp{\varepsilon}
\def\ep{\epsilon}
\def\ze{\zeta}
\def\io{\iota}
\def\ka{\kappa}
\def\la{\lambda}
\def\rh{\rho}
\def\na{\nabla}
\def\pa{\partial}
\def\ro{\varrho}
\def\si{\sigma}
\def\om{\omega}
\def\ph{\varphi}
\def\ta{\tau}
\def\th{\theta}
\def\te{\vartheta}
\def\up{\upsilon}
\def\Ga{\Gamma}
\def\De{\Delta}
\def\La{\Lambda}
\def\Si{\Sigma}
\def\Om{\Omega}
\def\Te{\Theta}
\def\Th{\Theta}

 \begin{center}

 {\Large \sc On the high derivative fermionic operator }
 \vskip 3mm
 {\Large \sc and trace anomaly}

 \vskip 10mm

 {\bf Guilherme de Berredo-Peixoto}\footnote{ E-mail address:
 peixoto@cbpf.br} \\ Centro Brasileiro de Pesquisas
 F\'{\i}sicas-CBPF-CNPq
 
\vskip 4mm

 and

\vskip 4mm
 {\bf Ilya L. Shapiro}\footnote{ 
E-mail address: shapiro@fisica.ufjf.br $\,\,,\,\,\,\,\,$
On leave from Tomsk Pedagogical  University, Tomsk, Russia. }
 \\
 Departamento de Fisica -- ICE, Universidade
 Federal de Juiz de Fora
 \\
 Juiz de Fora, 36036-330, MG, Brazil

 \end{center}
 \vskip 10mm


{\centerline{\large \it {\sl ABSTRACT}}}
\vskip 3mm
We construct a new example of the high derivative 
four-dimensional conformal
operator. This operator acts on fermions, and its
contribution to the trace anomaly has opposite sign, as
compared to conventional scalars, spinors and vectors.
Possible generalizations and applications are discussed.
\vskip 10mm

\section{Introduction}

The local conformal symmetry plays great role in
gravitational physics. But, the most important is that, 
due to the  quantum effects of the matter fields, 
conformal symmetry is violated by the trace anomaly 
\cite{duff}. This anomaly is relevant for the 
applications of quantum field theory on curved space-time 
\cite{birdav,duff94}. In particular, it is in the heart
of such achievement as the first inflationary cosmological 
model of Starobinsky \cite{star} and the semiclassical 
approach to the derivation of the Hawking radiation from the 
black holes \cite{black}. Recently, there was a considerable
interest to study the general properties of the anomaly
through the anomaly induced effective action \cite{reigert,frts}
(see also \cite{deser} for the consequent discussions). 
In particular, this effective action has been used to 
obtain systematic classification of the black hole vacuum
states \cite{balbi} and for the more detailed analysis of 
the Starobinsky model \cite{anju}. One has to notice, that the 
anomaly-induced action is defined with accuracy to an arbitrary 
conformal invariant functional \cite{buodsh}. Since, in general, 
there is no regular way to derive the conformal part of the 
effective action, it is useful to have
various versions of the conformal invariant actions, so that 
one could apply them to mimic the unknown conformal functional
\cite{balbi}. We remark, that the form of some conformal 
invariants constructed from curvature tensor has been already 
discussed my mathematicians (see, e.g. \cite{branson}).

The anomaly induced gravitational action is a direct $4d$ 
analog of the Polyakov action in $2d$. Since the quantum 
analysis of the Polyakov theory was led to the numerous 
interesting advances (starting from \cite{KPZ}), it is 
quite natural the emergence of the idea to use the $4d$ 
induced action of \cite{reigert,frts} to construct the
quantum theory of gravity \cite{odish,antmot} and to generalize 
the $c$-theorem from $2d$ for the $4d$ space-time 
\cite{cardy}. With respect to the renormalization group 
one can mention that the contribution of all matter fields
to anomaly and to the renormalization group equations 
for the parameter of the vacuum action possess some 
amusing universality. If we write the vacuum action in the
form 
\beq
S_{vacuum} \,=\, \int d^4x\sqrt{-g}\,
\left\{ a_1 C^2 +a_2 E + a_3 {\Box} R \right\}\,,
\label{vac}
\eeq
then all existing matter (non-gravitational) fields: scalars
\beq
S_0\,=\,\int d^4x\sqrt{-g}\,
\left\{\,\frac12\,g^{\mu\nu}\,\pa_\mu\phi\pa_\nu\phi +
\frac{1}{12}\,R\phi^2\,\right\}
\label{scal}
\eeq
spinors
$$
S_{1/2}\,=\,\frac{i}{2}\int d^4x\sqrt{-g}\,\left\{\,
\bar{\Psi}\ga^\mu\na_{\mu}\Psi
-\na_\mu\bar{\Psi}\,\ga^{\mu}\Psi\,\right\}\,=
$$
\beq
=\,i\int d^4x\sqrt{-g}\,{\bar \Psi}\ga^\mu\na_{\mu}\Psi
\,\,\,+\,\,\, {\rm surface\,\,\,\,\,\,\,\,term}
\label{spin}
\eeq
and vectors
\beq
S_{1} = \int d^4 x\sqrt{-g}\,
\Big\{\, - \frac14\,F_{\mu\nu}F^{\mu\nu}\,\Big\}
\label{vec}
\eeq
give positive contribution to the $\be$-function of the parameter $a_1$ 
and negative contribution to the $\be$-function of the parameter $a_2$
(see formulas (\ref{oef}) below). 
In relation to the beta-function of the $a_2$ parameter the situation 
remains the same in the framework of a supergravity theory. 
Besides (\ref{scal}), (\ref{spin}) and (\ref{vec}), conformal supergravity 
includes Weyl gravity (spin-2) and corresponding spin-$3/2$ fields. 
This universality has one important consequence: one can not cancel anomaly 
by choosing the appropriate number of the fields of different spins. 
Therefore, in this point one meets an important deviation between $2d$ 
and $4d$ cases, because in $2d$ the anomaly cancellation is indeed 
possible and this provides the existence of the critical dimension
in string theory. 

Taking into account the interest to the higher (than $2d$) dimensional
conformal field theories, one can formulate two relevant 
questions: i) whether one can construct the conformal invariant
theories distinct from the theories (\ref{scal}), (\ref{spin})
and (\ref{vec})? $\,\,\,\,$ ii) If this is possible, what would 
be the contribution to anomaly of these fields? In principle, it
might happen that including some special amount of these new 
fields into the definition of the integration measure, one could 
cancel the anomaly. In this case these new conformal fields 
would violate universality of the renormalization group flow. 
Summing up, the second question can be formulated as: whether it
is possible to maintain the conformal symmetry at quantum level 
by introducing some new fields? 

Once the investigation of the second problem has
been performed \cite{frts-sugra} and it was found that the 
cancellation is impossible for the fields which compose the 
conformal supergravity theory. As an example of the work done
in direction i), one can indicate Ref. \cite{erd}, where new tensor 
operators with local conformal symmetry have been constructed. 
On the other hand, it was noticed long ago \cite{pan,reigert,frts}
that there is the high derivative conformal operator
\beq
\De = \Box^2 + 2R^{\mu\nu}\,\na_\mu\na_\nu
-\frac23\,R\Box +\frac13\,(\na_\mu\,R)\,\na^\mu\,.
\label{paneitz}
\eeq
Moreover, the contribution of the corresponding free scalar 
\beq
S_4\,=\,\int d^4x\sqrt{-g}\,\ph\De\ph
\label{4sca}
\eeq
to the trace anomaly has opposite sign \cite{reigert}, 
as compared to the usual fields (\ref{scal}), (\ref{spin}) and 
(\ref{vec})\footnote{In \cite{reigert} the contribution of 
(\ref{paneitz}) 
to the anomaly was taken with negative sign, for this was taken as
the compensation of the integration over the auxiliary field.}.

Why the existence of two different conformal scalars (\ref{scal})
and (\ref{4sca}) is possible? The remarkable difference between 
two conformal scalars is the transformation law for the fields
\beq
g_{\mu\nu} \to g^\prime_{\mu\nu}=g_{\mu\nu}\,e^{2\si}
\,,\,\,\,\,\,\,\,\,\,\,\,\,\,\,
\phi \to \phi^\prime=\phi\,e^{-\si}
\,,\,\,\,\,\,\,\,\,\,\,\,\,\,\,
\ph \to \ph^\prime=\ph\,.
\label{transf}
\eeq

The main purpose of the present article is to construct the spinor
analog of the action (\ref{4sca}). The form of the first integral 
in (\ref{spin}) is dictated by the requirement of action to be 
Hermitian. Furthermore, in the conventional spinor case (\ref{spin})
the conformal transformation has the form
\beq
\Psi \to \Psi^\prime=\psi\,e^{-3\si/2}
\,,\,\,\,\,\,\,\,\,\,\,\,\,\,\,
{\bar \Psi} \to {\bar \Psi}^\prime={\bar \psi}\,e^{-3\si/2}
\label{transfer}
\eeq
Therefore, in order to construct the high derivative conformal 
action, one has to try the Hermitian action
\beq
S_3\,=\,\frac{i}{2}\,\int d^4x\sqrt{-g}\,\left\{\,
\bar{\psi}\ga^{\mu}{\cal D}_{\mu}\psi
-{\cal D}_{\mu}\bar{\psi}\,\ga^{\mu}\psi\,\right\}\,,
\label{acaofer}
\eeq
where $\,{\cal D}_\mu\,$ is some third derivative covariant 
operator, and postulate the following transformation law for 
the spinor $\psi$:
\beq
\psi \to \psi^\prime=\psi\,e^{-\si/2}
\,,\,\,\,\,\,\,\,\,\,\,\,\,\,\,
{\bar \psi} \to {\bar \psi}^\prime={\bar \psi}\,e^{-\si/2}\,.
\label{transferra}
\eeq
Indeed, these requirements guarantee the global conformal invariance
with $\,\si=const$ \cite{durr}. However, the possibility to have local 
conformal invariance is not at all obvious and we are going to 
investigate it here. 

The article is organized as follows. In section 2 we present the
construction of the conformal operator ${\cal D}_\mu$, and in
section 3 -- the calculation of its contribution to the trace anomaly.
In the last section we draw our conclusions and present some
mathematical {\it conjecture} and some speculations about
possible physical applications of this and other possible
conformal operators.

\section{The derivation of the conformal operator}   

The covariant derivative of the spinor is defined in a
usual way
$$
\na_{\mu}\psi =\partial_{\mu}\psi
+\frac{i}{2}\,\om^{ab}\mbox{}_{\mu}\Si _{ab}\psi\;,\;\;\;
\na_{\mu}\bar{\psi} =\partial_{\mu}\bar{\psi}
-\frac{i}{2}\,\om^{ab}\mbox{}_{\mu}\bar{\psi}\Si _{ab}\, ,
$$
where $\,\Si_{ab}=\frac{i}{2}\,(\ga_a\ga_b-\ga_b\ga_a)$. 
Consequently,
$$
[\na_{\mu}\, ,\na_{\nu}]\,\psi = \hat{{\cal R}}_{\mu\nu}\,\,\psi\, =\, 
\frac{1}{4}\,\ga^{\al}\ga^{\be}R_{\al\be\mu\nu}\,\psi\,.
$$

Using dimensional reasons, one can fix the possible form of the
operator $\,{\cal D}_{\mu}\,$ as
\beq
{\cal D}_{\mu}=\na_{\mu}\Box + k_1\,R_{\mu\nu}\na^{\nu}+k_2\,R\na_{\mu}\,,
\label{nachalo}
\eeq
where $\,k_1\,$ and $\,k_2\,$ are some unknown coefficients.
Integrating by parts and omitting the surface terms, we get
\beq
S\,=\,i\,\int d^4x\sqrt{-g}\,
\bar{\psi}\ga^{\mu}\tilde{{\cal D}}_{\mu}\psi +\, ...
\label{acaotildeD}
\eeq
where
\beq
\tilde{{\cal D}}_{\mu}=
\frac{1}{2}(\na_{\mu}\Box +\Box\na_{\mu})
+k_1R_{\mu\nu}\na ^{\nu}+k_2R\na_{\mu}
+\left( \frac{k_1}{4}+\frac{k_2}{2}\right) (\na_{\mu}R)\,.
\label{tildeDk}
\eeq
Making commutations of the covariant derivatives
\beq
\frac{1}{2}(\na _{\mu}\Box +\Box\na _{\mu})=
\na _{\mu}\Box +\frac{1}{2}[\Box\, ,\na _{\mu}]
\eeq
one can calculate
\beq
[\Box\, ,\na _{\mu}]\psi &
= & [\na _{\rho}\, ,\na _{\mu}]\na ^{\rho}\psi +
\na ^{\rho}[\na _{\rho}\, ,\na _{\mu}]\psi = \nonumber
\\
& = & -\frac{1}{2}\ga ^{\al}\ga ^{\be}R_{\al\be\mu\rho}\na^{\rho}\psi
+ R_{\mu\rho}\na ^{\rho}\psi
-\frac{1}{4}\ga^{\al}\ga^{\be}(\na^{\rho}R_{\al\be\mu\rho})\psi\,.
\label{comutbox}
\eeq
Substituting (\ref{comutbox}) into (\ref{tildeDk}), we obtain
the useful form of the operator.
\beq
\tilde{{\cal D}}_{\mu}\,=\,
\na _{\mu}\Box+a_1R_{\mu\rho}\na^{\rho}+a_2R\na_{\mu}
+a_3(\na_{\mu}R)\, ,
\label{tildeD}
\eeq
where $a_1=k_1$, $a_2=k_2$ e $a_3=\frac{a_1}{4}+
\frac{a_2}{2}-\frac{1}{8}$ -- the last is a condition of
Hermiticity. Our purpose will be to find such a values of
$a_{1,2,3}$, which provide both Hermiticity and conformal
invariance of the action (\ref{acaofer}).

For the one-parameter Lie group, one can safely restrict the
consideration by the infinitesimal version of the transformation
$$
g_{\mu\nu}\to g_{\mu\nu}^{'}=(1+2\si )g_{\mu\nu}
\, ,  \;\;\;\;\;\;\;\;
\psi\to\psi^{'}=(1-{\si}/{2})\psi
\,, \;\;\;\; \;\;\;\;
\bar{\psi}\to\bar{\psi}^{\prime}=(1-{\si}/{2})\bar{\psi}\, .
$$
Then, disregarding the highest orders in $\,\si$, after some
long algebra we arrive at the following transformations
\beq
(\bar{\psi}\ga ^{\mu}\na _{\mu}\Box\psi )^{'} & = &
-(\na _{\mu}\bar{\psi}\ga ^{\mu}\Box\psi )^{'}
+\na ^{'}_{\mu}(\bar{\psi}\ga ^{\mu}\Box\psi )^{'}
= \nonumber \\
& = & -(1-4\si )\na _{\mu}\bar{\psi}\ga ^{\mu}\Box\psi
-\na _{\mu}\si\bar{\psi}\ga ^{\mu}\Box\psi
-\na _{\mu}\bar{\psi}\ga ^{\mu}\na _{\nu}\si\na ^{\nu}\psi +\nonumber \\
& + & i\na _{\mu}\bar{\psi}\ga ^{\mu}\Si
_{\rho\nu}\na ^{\nu}\si\na ^{\rho}\psi
+\frac{1}{2}\na _{\mu}\bar{\psi}\ga ^{\mu}\Box\si\psi +
\na ^{'}_{\mu}(\bar{\psi}\ga ^{\mu}\Box\psi )^{'}\, ,
\eeq
\beq
(\bar{\psi}\ga ^{\mu}R_{\mu\rho}\na ^{\rho}\psi )^{'} = & &
(1-4\si )\bar{\psi}\ga ^{\mu}R_{\mu\rho}\na ^{\rho}\psi -\frac{i}{2}
\bar{\psi}\ga ^{\mu}R_{\mu}\mbox{}^{\nu}\Si _{\nu\rho}\na ^{\rho}\si\psi
- \nonumber \\
& - & \frac{1}{2}\bar{\psi}\ga ^{\mu}R_{\mu\nu}\na ^{\nu}\si\psi
- 2\bar{\psi}\ga ^{\mu}\na _{\mu}\na _{\nu}\si\na ^{\nu}\psi -
\nonumber \\
& - & \bar{\psi}\ga ^{\mu}\Box\si\na _{\mu}\psi\, ,
\eeq
\beq
(\bar{\psi}\ga ^{\mu}R\na _{\mu}\psi )^{'} =
(1-4\si )\bar{\psi}\ga ^{\mu}R\na _{\mu}\psi -
6\bar{\psi}\ga ^{\mu}\Box\si\na _{\mu}\psi +
\bar{\psi}\ga ^{\mu}R\na _{\mu}\si\psi \, ,
\eeq
\beq
(\bar{\psi}\ga ^{\mu}\na _{\mu}R\psi )^{'} =
(1-4\si )\bar{\psi}\ga ^{\mu}\na _{\mu}R\psi -
2\bar{\psi}\ga ^{\mu}R\na _{\mu}\si\psi -
6\bar{\psi}\ga ^{\mu}\na _{\mu}\Box\si\psi\, .
\eeq
Substituting these formulas into (\ref{acaotildeD})
with (\ref{tildeD}), we find that the conformal invariance
\beq
(\sqrt{-g}\,\bar{\psi}\ga ^{\mu}\tilde{{\cal D}}_{\mu}\psi )^{'}=
\sqrt{-g}\,\bar{\psi}\ga ^{\mu}\tilde{{\cal D}}_{\mu}\psi
\eeq
holds for the unique choice of the Hermitian parameters
\beq
a_1=1
\; ,\;\;\;\;\;\;\;\;\;\;\;\;\;\;\;\;\;\;
a_2=-\frac{5}{12}
\; ,\;\;\;\;\;\;\;\;\;\;\;\;\;\;\;\;\;\;
a_3=-\frac{1}{12}\, .
\label{coefs}
\eeq

\section{One-loop divergences and anomaly}

The one-loop effective action, for the free theory of field $\psi$
can be presented in the form
\beq
\Ga ^{(1)}=-i\Tr ln(\ga ^{\mu}\tilde{{\cal D}}_{\mu})=
-\frac{i}{2}\Tr ln(\ga^{\mu}\tilde{{\cal D}}_{\mu}
\ga^{\nu}\tilde{{\cal D}}_{\nu})\,,
\label{divercompact}
\eeq
that can be further rewritten as
\beq
\ga ^{\mu}\tilde{{\cal D}}_{\mu}\ga ^{\nu}\tilde{{\cal D}}_{\nu}=
\tilde{{\cal D}}_{\mu}\tilde{{\cal D}}^{\mu}+
\frac{1}{2}\ga ^{\mu}\ga ^{\nu}
\left[\tilde{{\cal D}}_{\mu}\, ,\, \tilde{{\cal D}}_{\nu}\right]\,.
\eeq

It proves useful to reduce the last expression to the 
form of the minimal six derivative operator
\beq
\ga^{\mu}\tilde{{\cal D}}_{\mu}
\ga^{\nu}\tilde{{\cal D}}_{\nu} = \hat{H} = {\hat 1}\,\Box^3
+\hat{V}^{\mu\nu}\na _{\mu}\na _{\nu}\Box
+ \hat{Q}^{\mu\nu\al}\na_{\mu}\na_{\nu}\na_{\al}
+\hat{U}^{\mu\nu}\na_{\mu}\na_{\nu}
+\hat{N}^{\mu}\na_{\mu} + \hat{P}\,.
\label{opercompact}
\eeq
By dimensional reasons, the $\hat{Q}^{\mu\nu\al},\,
\hat{N}^{\mu}\,$ and $\,\hat{P}$ - terms can not contribute
to the divergences, and therefore have no interest for us.
Indeed, this can be checked explicitly. 
The derivation of the divergences can be performed
using the generalized Schwinger-DeWitt technique developed in
\cite{bavi}. Since all the steps in this calculus are quite 
similar to the ones presented in \cite{bavi} for the four-derivative 
operator, we will not exhibit the details here. The general 
expression for the divergences has the form 
$$
-\frac{i}{2}\Tr ln\, \hat{H}\Big|_{div}
\,=\,-\, \frac{\mu^{n-4}}{(4\pi)^2\,(n-4)}\,\int d^nx\sqrt{-g}\,
\left\{\,
\frac{7}{120}\,R^2_{\mu\nu\al\be}+\frac{1}{15}\,R^2_{\mu\nu}
-\frac16\, R^2 -\frac25\,\Box R \right. +
$$
\beq
\left. + \tr\,\Big[ \frac12\hat{V}^{\mu\nu}\hat{{\cal R}}_{\mu\nu}+
\frac16\,\hat{V}^{\mu\nu}\,(R_{\mu\nu}-\frac12\,R\,g_{\mu\nu})
+ \frac14\,g^{\mu\nu}\,\hat{U}_{\mu\nu}
-\frac{1}{24}\,
\hat{V}^{\mu\nu}\hat{V}_{\mu\nu}-\frac{1}{48}\,(\hat{V}^{\mu\nu}g_{\mu\nu})^2
\,\Big]
\,\right\}\,,
\label{BV}
\eeq
and of course they do not depend on $\hat{Q}^{\mu\nu\al},\,
\hat{N}^{\mu}\,$ and $\,\hat{P}$ - terms. 

Now we are in a position to calculate the divergent part of the
one-loop effective action using (\ref{BV}). For the sake of 
generality we shall perform the calculations for arbitrary 
values of $a_{1,2,3}$, and will substitute (\ref{coefs}) only
afterwards. 
After some long calculus, disregarding the non-essential 
terms with more than four derivatives of the external metric, 
we arrive at the relations
\beq
\ga^{\mu}\ga^{\nu}
\left[\tilde{{\cal D}}_{\mu}\, ,\, \tilde{{\cal D}}_{\nu}\right] & = &
\frac{1}{8}\ga^{\al}\ga^{\be}\ga^{\mu}\ga^{\nu}
R_{\al\be\rho\la}R_{\mu\nu}\mbox{}^{\rho\la}\Box -
2\ga^{\mu}\ga^{\nu}R_{\nu\rho}\na _{\mu}\na^{\rho}\Box -\nonumber \\
& - &\frac{1}{2}\Box R\Box - \frac{1}{2}R\Box^2 +
[\ga^{\mu},\ga^{\nu}]R_{\nu\rho}\na _{\mu}\na^{\rho}\Box +
\nonumber \\
& + & a_1\ga ^{\mu}\ga ^{\nu}\ga ^{\al}\ga ^{\be}\left(
-R_{\al\be\si\rho}R^{\si}\mbox{}_{[\nu}\na _{\mu ]}\na ^{\rho}+
\frac{1}{2}R_{\al\be [\mu}\mbox{}^{\si}R_{\nu ]\si}\Box\right) +
\nonumber \\
& + &
a_2R^2\Box -2a_2\ga ^{\mu}\ga ^{\nu}RR_{\nu\rho}\na _{\mu}\na ^{\rho}
+\, ... \\
\tilde{{\cal D}}_{\mu}\tilde{{\cal D}}^{\mu} & = &
\Box ^3 +(2a_1+1)R_{\mu\nu}\na ^{\mu}\na ^{\nu}\Box -\frac{1}{16}
\ga ^{\al}\ga ^{\be}\ga ^{\la}\ga ^{\si}
R_{\al\be\mu\rho}R_{\la\si}\mbox{}^{\mu\rho}\Box +
\nonumber \\
& + & a_1\Box R^{\mu\nu}\na _{\mu}\na _{\nu}+
2a_1\na _{\mu}\na _{\rho}R^{\mu\si}\na ^{\rho}\na _{\si} +
\nonumber \\
& + & (a_1+a_1^2)R^{\mu\si}R_{\si\rho}\na _{\mu}\na ^{\rho}
+(a_2+a_3)\Box R\Box + \nonumber \\
& + & 2(a_2+a_3)\na _{\mu}\na _{\nu}R\na ^{\mu}\na ^{\nu}+
(a_2+2a_1a_2)RR_{\mu\nu}\na ^{\mu}\na ^{\nu} +
\nonumber \\
& + & 2a_2R\Box ^2 +a_2^2R^2\Box +\, ...\, .
\eeq
Then, the relevant blocks of (\ref{opercompact}) are
\beq
\hat{V}^{\mu\nu} & = & 2a_1R^{\mu\nu}+(2a_2-\frac{1}{4})Rg^{\mu\nu} \\
\hat{U}^{\mu\nu} & = &
a_1\Box R^{\mu\nu}+2a_1\na _{\rho}\na ^{\mu}R^{\rho\nu}+
(a_2+a_3-\frac{1}{4})g^{\mu\nu}\Box R+ \nonumber \\
& + & (a_1^2+a_1)R^{\mu}\mbox{}_{\rh}R^{\rh\nu}+
2(a_2+a_3)\na ^{\mu}\na ^{\nu}R+
(2a_1a_2+a_2)RR^{\mu\nu}+ \nonumber \\
& + & (a_1^2-\frac{a_2}{2})g^{\mu\nu}R^2 -
\frac{1}{2}a_1\ga ^{[\mu}\ga ^{\rh ]}\ga ^{\al}\ga ^{\be}
R_{\al\be}\mbox{}^{\si\nu}R_{\si\rh}
-\frac{a_1}{2}R_{\al\be}^2g^{\mu\nu}-
\nonumber \\
& - & a_2\ga ^{\mu}\ga ^{\rh}R_{\rh}\mbox{}^{\nu}R\, .
\eeq

Using the formula (\ref{BV}), for the generic  
operator (\ref{tildeD}) we arrive at the following divergences
\beq
\Ga ^{(1)}_{{\rm div}}=
-\frac{1}{\vp}\int d^4x\sqrt{-g}\,\left\{\,
\al R_{\mu\nu\al\be}^2 + \be R_{\mu\nu}^2 + \ga R^2 +\de \Box R
\,\right\}\, ,
\eeq
where
\beq
\al & = & \frac{7}{120}\; ,\;\;\;\;\;\;
\be =\frac{1}{3}a_1^2-\frac{2}{3}a_1+\frac{1}{15}\, , \\
\nonumber
\ga & = & -\frac{1}{3}a_1^2-4a_2^2-2a_1a_2-\frac{1}{6}a_1-\frac{4}{3}a_2-
\frac{1}{8}\, , \\
\nonumber
\de & = & 2a_1 +6a_2 +6a_3-\frac{7}{5}\, .
\label{valoresalfa}
\eeq
Replacing the coefficients (\ref{coefs}) corresponding to the
Hermitian conformal operator, and using the basis of the square
of the Weyl tensor, Gauss-Bonnet term
$$
C^2=C_{\mu\nu\al\be}C^{\mu\nu\al\be} =
R_{\mu\nu\al\be}^2-2R_{\mu\nu}^2+\frac{1}{3}R^2\,,
\,\,\,\,\,\,\,\,\,\,\,\,\,
E=R^2_{\mu\nu\al\be}-4R_{\mu\nu}^2+R^2
$$
and $\,R^2$\,,
we arrive at the final result
\beq
\Ga^{(1)}_{{\rm div}}\,=\,-\,
\frac{\mu^{(n-4)}}{\vp}\int d^nx\sqrt{-g}\,\left\{\,
-\frac{1}{60}C^2+\frac{3}{40}E-\frac{12}{5}\Box R\,\right\}\,,
\label{resfinaldiv}
\eeq
which is conformal invariant up to total derivatives. The
cancellation of the non-conformal $\sqrt{-g}R^2$-term confirms
the correctness of our calculations of both conformal operator
and divergences.

The conformal anomaly is directly related to the divergences
\cite{duff}, so that we have, using (\ref{resfinaldiv}):
\beq
<T^{\mu}\mbox{}_{\mu}>=-
\frac{1}{2}\frac{1}{\sqrt{-g}}\frac{\de \De S}{\de\si}=
-\frac{1}{(4\pi )^2}\,\left( \om C^2+b E+c\Box R\right)\, .
\label{anomconfgeral}
\eeq
with
\beq
\om=-\frac{1}{60}
\,,\,\,\,\,\,\,\,\,\,\,\,\,\,\,\,\,\,\,\,\,
b=+\frac{3}{40}
\,,\,\,\,\,\,\,\,\,\,\,\,\,\,\,\,\,\,\,\,\,
c=-\frac{12}{5}
\label{coefi}
\eeq

Consider the cancellation of anomaly. In the four-dimensional space 
($D=4$), if one has only conventional scalar, spinor and vector 
fields, the cancellation of anomaly is impossible due to the fact that 
all these fields contribute to the coefficients $\om$ and $b$
(\ref{coefi}) with the same signs.
But, situation might change if we have some high derivative
fields. For instance, some examples of finite and anomaly-free 
theory has been given in \cite{eli}, where the IR and UV conformal 
fixed points of the renormalization group flow were established.
Now, we shall see, that including the new conformal fields one can 
achieve the anomaly cancellation in a different way. 

Imagine that we have a theory with $N_0$ real massless
conformal invariant scalars (\ref{scal}),
$N_{1/2}$ Dirac spinors  (\ref{spin}) and $N_1$ massless
vectors. In addition, the theory includes $n_3$ copies of the
high derivative spinor (\ref{acaofer}) and $n_4$ copies
of the high derivative scalar (\ref{4sca}). Then the total
expression for the anomaly is (\ref{anomconfgeral}), with the
following total coefficients:
$$
\om_t \,=\, \left(\,\frac{1}{120}\,N_0 + \frac{1}{20}\,N_{1/2} 
+ \frac{1}{10}\,N_1\,\right)
\,-\, \left(\,\frac{1}{15}\,n_4 + \frac{1}{60}\,n_3\,\right)\,,
$$
$$
b_t\,=\,-\,\left(\,\frac{1}{360}\,N_0 + \frac{11}{360}\,N_{1/2}
+ \frac{31}{180}\,N_1\,\right) 
\,+ \,\left(\,\frac{7}{90}\,n_4  + \frac{3}{40}\,n_3\,\right)\,,
$$
\beq
c_t=\frac{N_0}{180} + \frac{N_{1/2}}{30} - \frac{N_1}{10}
-\frac{2}{45}\,n_4-\frac{12}{5}\,n_3\,.
\label{oef}
\eeq
In order to obtain the conditions of anomaly cancellation,
one has to consider only the coefficients
$\om_t$ and $b_t$, because $c_t$ can be always canceled by adding the
local finite counterterm \cite{duff}
\beq
\De S_c \,=\, \frac{c_t}{12\,(4\pi)^2}\,\int d^4x\sqrt{-g}\,R^2\,.
\label{cont-de}
\eeq
Then, from (\ref{oef}) one arrives at the following solutions
for the amount of the high derivative fields
\beq
n_3 = N_1 - \frac12\,N_{1/2} - \frac18\,N_{0}
\,,\,\,\,\,\,\,\,\,\,\,\,\,\,\,\,\,\,\,\,\,\,
n_4 = \frac{5}{4}\,N_1 + \frac{7}{8}\,N_{1/2}
+\frac{5}{32}\,N_{0} \,.
\label{netano}
\eeq
Indeed, both $n_3$ and $n_4$ must be integers.
One can see, that the cancellation of trace anomaly is, in principle,
possible, but
it puts some restrictions on the field composition $N_{0,1/2,1}$
of the matter theory. Namely: we need 
$\,\, N_0+4N_{1/2}\leq 8N_1 $,
$N_0$ to be multiple of 32, $N_{1/2}$ to be multiple of 8
and  $N_1$ to be multiple of 4. 
The last conditions can be easily 
satisfied for some gauge groups. Then, in some theory with extended
supersymmetry, if the one-loop anomaly is exact, the proper
choice of the numbers $n_3$ and $n_4$ can provide the complete
cancelation of anomaly.

\section{Conclusions and speculations}

We have constructed the 3-derivative spinor action which possesses
local conformal invariance. Our solution is a generalization
of the flat-space operators \cite{durr} with global scale
invariance. The relation between the corresponding
spinor and conventional massless Dirac spinor is similar to the
one between fourth derivative scalar (\ref{4sca})
and usual conformal scalar (\ref{scal}). One can formulate
the following mathematical {\it conjecture}: those are only
first representatives of the infinite family of the conformal
invariant operators with even (for scalars) and odd (for spinors)
number of derivatives in $D=4$. The scalars and spinors corresponding
to these operators transform according to their
classical dimension. The generalization to the $D\neq 4$ is
also possible (see \cite{alter} for the $\De$ operator). To check
this {\it conjecture} would be an interesting mathematical problem.

The contributions of a new third derivative spinor to the trace
anomaly has the sign opposite to the one of the usual scalars,
spinors and vectors, and the same as for the high derivative scalar
(\ref{4sca}). One can guess that this sign distribution is
related to the emergence and dominating contributions of the
high derivative unphysical ghosts, which are always present in
the spectrum of the high derivative operators.
As an extension of our {\it conjecture}, one can suppose
that the signs of the contributions of the (yet unknown)
higher order conformal
operators to the trace anomaly coefficients $\om_t$ and $b_t$
will alter, as in (\ref{oef}). Therefore, if there exists a
SYM theory with extended supersymmetry, for which the one-loop
anomaly is exact, one can define the integration measure, in
curved space-time, in such a way that the $D=4$ conformal
symmetry is exact. This definition of the measure must include
proper number of functional determinants of the operators
like $\,\De\,$ and $\,\gamma^\mu\tilde{\cal D}_\mu$. It might happen,
that the investigation of this hypothesis can shed some light
on the supersymmetry breaking mechanism which may be, after
all, related to the conformal anomaly.

\vskip 6mm\vskip 6mm

\noindent
{\bf Acknowledgments.}
\vskip 3mm

We are grateful to I.L. Buchbinder for useful discussion, 
and to the CNPq (Brazil) for grant (I.Sh.) and schoolarship 
(G.B.P.).  I.Sh. is also grateful to the RFFI (Russia) for 
the support of the group 
of theoretical physics at Tomsk Pedagogical University
through the project 99-02-16617.



\begin{thebibliography}{99}

\bibitem{duff} M.J. Duff, 
Nucl. Phys. {\bf 125 B} (1977) 334.

\bibitem{birdav} N.D. Birell and P.C.W. Davies, {\sl Quantum fields
in curved space} (Cambridge Univ. Press, Cambridge, 1982).

\bibitem{duff94} M.J. Duff, Class. Quant. Grav. {\bf 11} (1994) 1387.

\bibitem{star}
A.A. Starobinski, Phys.Lett. {\bf 91B} (1980) 99.

\bibitem{black} S.M. Christensen and S.A. Fulling, 
{\sl Phys. Rev.} {\bf D15} (1977) 2088.

\bibitem{reigert} R.J. Riegert, 
Phys. Lett. {\bf B 134} (1984) 56.

\bibitem{frts} E.S. Fradkin and A.A. Tseytlin,
Phys. Lett. {\bf B 134} n.3,4 (1984) 187.

\bibitem{deser}
S. Deser and A. Schwimmer, Phys.Lett. {\bf 309B} (1993) 279;

S. Deser, Phys.Lett. {\bf 479B} (2000) 315.

\bibitem{balbi} R. Balbinot, A. Fabbri and I.L. Shapiro,
Nucl.Phys. {\bf B559} (1999) 301. 

\bibitem{anju} J.C. Fabris, A.M. Pelinson, I.L. Shapiro,
{\sl On the gravitational waves on the background 
of anomaly-induced inflation.} Nucl. Phys. {\bf B}, 
to be published.

\bibitem{buodsh} I.L. Buchbinder, S.D. Odintsov and I.L. Shapiro,
Phys.Lett. {\bf 162B} (1985) 92.

\bibitem{branson} 
T.P. Branson, Comm. Partial Diff. Equations {\bf 7} (1983) 305;
Comm. Math. Phys. {\bf 178} (1996) 301.

\bibitem{KPZ} 
V.G. Kniznik, A.M. Polyakov and A.B. Zamolodchikov,
Mod. Phys. Lett. {\bf 3A} (1988) 819.

\bibitem{odish} S.D. Odintsov and I.L. Shapiro,
Class. Quant. Gravity. {\bf 8} (1991) L57.

\bibitem{antmot} I. Antoniadis and E. Mottola,
Phys. Rev. {\bf 45D}, 2013 (1992).

\bibitem{cardy} J.L. Cardy, Phys. Lett. {\bf 215B} (1988) 749;

I. Jack and H. Osborn, Nucl. Phys. {\bf B343} (1990) 647;

A. Capelli, D. Friedan and  J.I. Latorre,  Nucl. Phys. {\bf B352} (1991) 616;

D.Z. Freedman and H. Osborn, Phys. Lett. {\bf 432B} (1998) 353.

\bibitem{frts-sugra} E.S. Fradkin and A.A. Tseytlin, Phys.Repts. 
{\bf 119} (1985) 233.

\bibitem{erd} 
J.  Erdmenger, Class. Quant. Grav. {\bf 14} (1997) 2061;

J.  Erdmenger and H. Osborn, Class. Quant. Grav. {\bf 15} (1998) 273.

\bibitem{pan}  S. Paneitz, A Quartic Conformally Covariant Differential
Operator for Arbitrary Pseudo-Riemannian Manifolds, MIT preprint, 1983
(unpublished).

\bibitem{durr} H.P. D\"urr and P. du T. van der Merwe,
Il Nuovo Cim. {\bf 23 A,n.1} (1974) 1;

C.C. Chiang and H.P. D\"urr, 
Il Nuovo Cim. {\bf 28 A,n.1} (1975) 89.

\bibitem{bavi} A.O. Barvinsky and G.A. Vilkovisky,
Phys. Repts. {\bf 119} (1985) 1.

\bibitem{eli} E.Elizalde, A.G.Jacksenaev, S.D.Odintsov and I.L.Shapiro,
Class.Quant.Grav. {\bf 12} (1995) 1385.

\bibitem{alter}  J.A. de Barros and  I.L. Shapiro,
Phys. Lett. {\bf 412B} (1997) 242.

\end{thebibliography}
\end{document}